\documentclass[%
superscriptaddress,
 preprint,
 amsmath,amssymb,
 aps,
 prl,
 letterpaper,
]{revtex4-2}

\usepackage{graphicx}
\usepackage{dcolumn}
\usepackage{bm}

\newcolumntype{.}{D{.}{.}{8}}

\usepackage[utf8]{inputenc}
\usepackage{physics}
\usepackage{amsmath, amssymb}
\usepackage{tabularx,booktabs,array,dcolumn}
\usepackage{setspace}
\usepackage{vmargin}
\usepackage{xcolor}
\usepackage{graphicx,psfrag,subfigure}

\usepackage{float}
\usepackage[all]{xy}

\newcommand{\mx}[1]{\mathbf{#1}}

\newcommand{\mr}[1]{\mathrm{#1}}
\newcommand{\pd}[2]{\frac{\partial #1}{\partial #2}}
\newcommand{\cm}{cm$^{-1}$}

\newcommand{\colred}{\color{red}}

\def\Eh{$E_\mathrm{h}$}

\def\m0{m^{(0)}}

\def\mnuc{m_\mr{nuc}}

\def\mel{m_\mr{e}}
\def\iim{i}

\def\m{\mathbf{m}}

\def\el{\mathrm{el}}

\def\he2p{He$_2^+$}
\def\h2p{H$_2^+$}

\newcommand{\epsi}{\varepsilon}
\def\mmx{\mathcal{M}}

\begin{document}

\title{%
Non-adiabatic, relativistic, and leading-order QED corrections 
for rovibrational intervals of $^4$He$_2^+$ ($X\ ^2\Sigma_\text{u}^+$)
}

\author{D\'avid Ferenc} 
\affiliation{Institute of Chemistry, ELTE, Eötvös Loránd University, Pázmány Péter sétány 1/A, Budapest, H-1117, Hungary}

\author{Vladimir I. Korobov}
\email{korobov@theor.jinr.ru}
\affiliation{Bogoliubov Laboratory of Theoretical Physics, Joint Institute for Nuclear Research, Dubna 141980, Russia}

\author{Edit M\'atyus}
\email{matyus@chem.elte.hu}
\affiliation{Institute of Chemistry, ELTE, Eötvös Loránd University, Pázmány Péter sétány 1/A, Budapest, H-1117, Hungary}

\date{\today}

\begin{abstract}
\noindent %
The rovibrational intervals of the $^4$He$_2^+$ molecular ion in its $X\ ^2\Sigma_\text{u}^+$ ground electronic state are computed by including the non-adiabatic, relativistic, and leading-order quantum-electrodynamics corrections. 
Good agreement of theory and experiment is observed for the rotational excitation
series of the vibrational ground state and the fundamental vibration.
The lowest-energy rotational interval is computed to be
$70.937\ 69(10)$~\cm\ 
in agreement with the most recently reported experimental value, 
$70.937\ 589(23)(60)_\text{sys}$~\cm\ 
[L. Semeria, P. Jansen, G.-M. Camenisch, F. Mellini, H. Schmutz, and F. Merkt, Phys. Rev. Lett. 124, 213001 (2020)].
\end{abstract}

\maketitle
\clearpage
\section{Introduction}
\noindent%
Few-electron molecules serve as benchmark systems for experimental and theoretical molecular physics and spectroscopy. 
Recent experimental and theoretical progress of
\h2p, H$_2$, and their isotopologues \cite{HoBeSaEiUbJuMe19,PuKoCzPa19,Korobov2020} 
is connected to proposals to test 
fundamental interactions \cite{AlDrSaUbEi18,rmp18}
and to refine fundamental physical constants \cite{KaHiKoKo16,AlHaKoSc18}
using molecular spectroscopy. 

The present work joins this direction and focuses on the five-particle $^4$He$_2^+$ molecular ion
in its ground electronic state ($X\ ^2\Sigma_\text{u}^+$). 
In addition to testing fundamental aspects, 
precision spectroscopy of $^4$He$_2^+$ in combination with accurate \emph{ab initio}
computations has been proposed as an alternative way to determine 
the polarizability of the helium atom \cite{PuPiKoJeSz16,SeJaCaMeScMe20}. 
Precise knowledge of this quantity is necessary for
a possible new definition of the pressure standard 
based on counting the number density of a sample of helium gas.
There has been experimental progress in the precision spectroscopy of $^4$He$_2^+$ 
including the measurement of  
the spin-rotational fine structure \cite{JaSeMe18prl} and
the rotational and rovibrational intervals \cite{JaSeMe16b,SeJaMe16,JaSeMe18,SeJaCaMeScMe20}.

The present work is concerned with the rotational and rovibrational intervals for which
disagreement was observed between the experimental results \cite{JaSeMe16b,SeJaMe16,JaSeMe18,SeJaCaMeScMe20}
and (lower-level) theoretical work \cite{TuPaAd12,Ma18he2p}.
The experimental `dataset' includes the rotational intervals for the vibrational ground state, $(0,N^+)$--$(0,1)\ (N^+=3,\ldots,19)$ \cite{SeJaMe16} and 
the rovibrational intervals connecting the ground and the first excited 
vibrational state $(1,N^+)$--$(0,1)\ (N^+=1,\ldots,13)$ \cite{JaSeMe18}
with an experimental uncertainty of $0.000\,8$~\cm\ and $0.001\,2$~\cm, respectively.
The lowest-energy rotational interval
is known more precisely to be $70.937\ 589(23)\pm 0.000\ 060_\text{sys}$~\cm\ \cite{SeJaCaMeScMe20}.
%

The most precise theoretical results for molecules can be obtained by 
including all electrons and nuclei in the non-relativistic quantum mechanical treatment  \cite{chemrev13,Ko18h2p,WaYa18,PuKoCzPa19,Ma19review,FeMa19EF}.
All bound rovibrational and several resonance states of H$_2^+$ treated as a three-particle system have been converged with an uncertainty in their non-relativistic 
energy better than 10$^{-7}$~\cm\ \cite{Ko18h2p}, 
and a similar precision has been achieved for selected states of H$_2$ treated 
as a four-particle system \cite{PaKo18four}.
The fundamental vibration energy has been computed for 
$^3$He$^4$He$^+$ treated as a five-particle system \cite{StBuAd09}, 
but the convergence error of 
this energy appears to be at least two-orders of magnitude larger than 
the uncertainty of the currently available experimental value of the parent isotopologue.

To ensure a direct comparison with the experimental dataset, which includes high
rotational angular momentum quantum numbers up to $N^+=19$, 
and a tight control of the numerical (convergence) error, 
we start out from the Born--Oppenheimer approximation 
and account for non-adiabatic corrections by perturbation theory \cite{Te03,PaSpTe07,PaKo09,Ma18nonad,MaTe19}. 
The experimental dataset belongs to the ground ($X\ ^2\Sigma_\text{u}^+$) electronic
state that is well-separated from the electronically excited
states over the relevant nuclear configuration range, hence we may expect
non-adiabatic perturbation theory to perform well.
%

There is some evidence about the increasing importance of the non-adiabatic effects
with rotational excitation of $^4$He$_2^+$ \cite{SeJaMe16,Ma18he2p}, but the non-adiabatic non-relativistic
computation of Ref.~\cite{Ma18he2p} was only partially able to account for 
the discrepancy between theory and experiment for the rotational series. 
Furthermore, the non-adiabatic corrections (without 
relativistic and QED effects) increased the deviation of theory and experiment for the fundamental vibration energy \cite{Ma18he2p,JaSeMe18} in comparison to the adiabatic result \cite{TuPaAd12}.

The present work reports a more complete theoretical treatment
for the rotational-vibrational intervals of $^4$He$_2^+$ ($X\ ^2\Sigma_\text{u}^+)$, and
we account for 
the non-adiabatic, relativistic and
leading-order QED corrections.
The error balance of the computational procedure is 
analyzed and further contributions, neglected in this work, are discussed.

First, we solved the electronic Schrödinger equation for $n=3$ electrons and $N=2$ fixed nuclei 
for the $\phi_0$ ground electronic state (in Hartree atomic units) 
\begin{align}
  &H_\el \phi_0(\mx{r},\mx{R}) 
  =
  E_{\text{el},0}(\mx{R})
  \phi_0(\mx{r},\mx{R}) \quad \text{with}\nonumber \\
  &H_\el 
  =
    -\sum_{i=1}^n \frac{1}{2\mel} \Delta_{\mx{r}_i}
    +\sum_{i=1}^n \sum_{j>i}^n \frac{1}{|\mx{r}_i-\mx{r}_j|}
    +\sum_{i=1}^n \sum_{j=1}^N \frac{Z_j}{|\mx{r}_i-\mx{R}_j|} \;
\end{align}
using floating explicitly correlated Gaussian (fECG) basis functions
and the QUANTEN computer program \cite{Ma19review,Ma18nonad}.

The rovibrational Hamiltonian corresponding to the ground electronic (`0'th) state
and accounting for non-adiabatic coupling up to the second-order terms in $\epsi=(\mel/\mnuc)^{\frac{1}{2}}$ ,  
is \cite{Te03,PaSpTe07,MaTe19} 
\begin{align}
  H^{(2)}_{0}
  =
  \sum_{i,j=1}^{3N}
    \frac{1}{2}
    (-\iim\epsi \partial_{R_i})
    (\delta_{ij}-\epsi^2 M_{ij})
    (-\iim\epsi \partial_{R_i})
    +
    E_{\el,0}
    +
    \epsi^2U_0 \; ,
  \label{eq:nadHam}
\end{align}
where
\begin{align}
  U
  =
  \frac{1}{2}
  \sum_{i=1}^{3N}
    \langle %
      \partial_{R_i} \phi_0
      |
      \partial_{R_i} \phi_0      
    \rangle
\end{align}
and 
\begin{align}
  M_{ij}
  =
  2\langle%
    \partial_{R_j} \phi_0 
    | P^\perp_0(\hat{H}_\text{e}-E_{\el,0})^{-1}P^\perp_0|
    \partial_{R_i}\phi_0
  \rangle
  \; ,
  \quad 
  P^\perp_0=1-|\phi_0\rangle\langle\phi_0| \; ,
\end{align}
are the diagonal Born--Oppenheimer correction (DBOC) and the mass-correction tensor,
respectively.
%

Rotational-vibrational states of He$_2^+$ are computed using this Hamiltonian written 
in spherical polar coordinates, $(\rho,\theta,\phi)$, which leads to
the solution of the radial equation \cite{Ma18nonad,Ma18he2p}:
\begin{align}
  &\left(%
    -\pd{}{\rho} 
    \frac{1}{\mnuc}
    \left[%
      1-\frac{\mmx^\rho\,_\rho}{\mnuc}
    \right] 
    \pd{}{\rho}
  \right.
  \nonumber \\
  &\left. 
   +\frac{N^+(N^++1)}{\rho^2}
    \frac{1}{\mnuc}
    \left[%
      1 - \frac{\mmx^\Omega\,_\Omega}{\mnuc}
    \right]
    +
    U(\rho) + E_\text{el}(\rho)
  \right)
  \chi_{N^+}(\rho) 
  =
  E_{N^+} \chi_{N^+}(\rho)  \; .
  \label{eq:diatrad2}
\end{align}
$\mmx^\rho\,_\rho$ and $\mmx^\Omega\,_\Omega$ are the vibrational and rotational mass correction functions corresponding to the curvilinear representation \cite{Ma18nonad}. 
The equation is solved for each $N^+$ rotational angular momentum quantum number 
using a discrete variable representation \cite{LiCa00}.\\

We have computed the $E_\text{el}(\rho)$ potential energy curve over the $\rho\in[0.992,3.5]$~bohr interval of the internuclear separation that is necessary to converge
the rovibrational states considered in this work.
As a result, the electronic energy at the equilibrium structure 
($\rho_\text{eq}=2.042$~bohr) is within the 0.2~$\mu$\Eh\ error bar of the complete basis set limit estimate by Cencek et al. \cite{CeRy00}.
The newly computed part of the potential energy curve (PEC) improves  
the earlier PEC \cite{TuPaAd12} by 0.012~\cm\ (59~n\Eh) at the equilibrium structure 
and by 0.034~\cm\ (155~n\Eh) at $\rho=3.5$~bohr.

Table~\ref{tab:err} collects the calculated change in the energy intervals using the newly computed and the earlier curves. 
As a (conservative) estimate for the remaining error due to uncertainties of the PEC, 
we used the half of the observed change.
We think that the 
uncertainty of the rovibrational intervals due to the uncertainty
of the PEC is within a few nE$_\text{h}$.

\begin{table}
  \caption{%
    Error balance of the rotational and (ro)vibrational intervals, in \cm, 
    computed in the present work. 
    The numerical uncertainty of the computed intervals is estimated based on the difference 
    in the intervals obtained with two different datasets:
    $^\text{a}$ PEC (DBOC) curve from Ref.~\cite{TuPaAd12} and from the present work;
    $^\text{b}$ non-adiabatic mass computed in Ref.~\cite{Ma18he2p} and in the present work;
    $^\text{c}$ relativistic corrections obtained with the integral transformation technique \cite{PaCeKo05} and the `direct' method \cite{StPaAd16};
    $^\text{d}$ effect of a hypothetical $\pm$1\% change in $\beta_\text{el}$;
    $^\text{e}$ $\sigma$ is obtained as the sum of the absolute value of the terms;
    $^\text{f}$ the effect of the neglected higher-order QED corrections is estimated with
     the dominant term of $E^{(4)}$, Eq.~(\ref{eq:hQED});
    $^\text{g}$ estimate for the coupling of the non-adiabatic and relativistic corrections, 
    see also Ref.~\cite{CzPuKoPa18};
    $^\text{h}$ estimate for the effect of the finite nuclear size \cite{Pachucki_H2_2009};    
    $^\ast$ We use half of this value for the uncertainty estimate of the present results.
    \label{tab:err}
  }
\begin{tabular}{@{}l D{.}{.}{7} D{.}{.}{6} c@{\ \ \ \ } D{.}{.}{5} D{.}{.}{5} @{}}
\cline{1-6}\\[-0.4cm] 
\cline{1-6}\\[-0.4cm]
 & \multicolumn{2}{c}{Rotational intervals} 
 && \multicolumn{2}{c}{(Ro)vibrational intervals} \\
 \cline{2-3}\cline{5-6}\\[-0.4cm] 
 & \multicolumn{1}{c}{(0,3)--(0,1)}
 & \multicolumn{1}{c}{$\mathrm{RMSD}_\mathrm{rot}$} 
 && \multicolumn{1}{c}{~~(1,0)--(0,0)} 
 & \multicolumn{1}{c}{~~~$\mathrm{RMSD}_\mathrm{rv}$}  \\ 
\cline{1-6}\\[-0.4cm] 
\multicolumn{6}{l}{\it Numerical uncertainty estimate for the computed terms ($\pm\sigma$):} \\
PEC$^\text{a}$ &	-0.000\ 002	&	0.000\ 15	&&	-0.003\ 28^\ast	&	0.003\ 37^\ast \\
DBOC$^\text{a}$ &	-0.000\ 010	&	0.000\ 18	&&	-0.000\ 16	&	0.000\ 19	\\
Nadm$^\text{b}$ &	-0.000\ 018	&	0.000\ 36	&&	-0.000\ 13	&	0.000\ 24	\\
Rel.$^\text{c}$ &	-0.000\ 012	&	0.000\ 18	&&	0.001\ 09	&	0.000\ 84	\\
$\beta_\text{el}$ ($\pm 1$\%)$^\text{d}$ &	-0.000\ 032	&	0.000\ 22	&&	0.000\ 12	&	0.000\ 63	\\
\cline{1-6}\\[-0.4cm] 
$\pm\sigma$$^\text{e}$ &	\pm 0.000\ 073	& \pm 0.001\ 09	&&	\pm 0.003\ 14	& \pm 0.003\ 59	\\[0.25cm]
%
\cline{1-6}\\[-0.4cm] 
\multicolumn{6}{l}{\it Estimate for neglected theoretical terms ($\Delta_\mathrm{est}$):} \\
hQED$^\text{f}$	    &	-0.000\ 008	&  -0.000\ 15	&&	-0.000\ 13	& -0.000\ 25	\\
Nad\&Rel$^\text{g}$	&	-0.000\ 001	&  -0.000\ 02	&&	-0.000\ 01	& -0.000\ 03	\\
Fsn$^\text{h}$	    &	-0.000\ 001	&  -0.000\ 02	&&	-0.000\ 02	& -0.000\ 04	\\
\cline{1-6}\\[-0.4cm] 
$\Delta_\text{est}$ &	-0.000\ 010	&  -0.000\ 19	&&	-0.000\ 16	& -0.000\ 31	\\
\cline{1-6}\\[-0.4cm] 
\cline{1-6}\\[-0.4cm]
\end{tabular}
\end{table}

The relativistic effects on the electronic motion
are accounted for by incrementing the $E_\text{el}+U$ adiabatic potential
energy curve with the expectation value of the spin-independent part of the Breit--Pauli Hamiltonian, including the mass-velocity term, the Darwin terms and the spin-spin coupling, 
as well as the orbit-orbit term \cite{BeSabook77}:
\begin{align}
E_{\mathrm{rel}}^{(2)}
=
\alpha^2
\matrixel{\phi_0}{H_\mathrm{rel}^{(2)}}{\phi_0} \ ,
\end{align}
where
\begin{align}
H_{\mathrm{rel}}^{(2)}
=
-&\frac{1}{8} \sum_{i=1}^n \mx{p}_i^4  
+ 
\frac{\pi}{2} 
\sum_{i=1}^n\sum_{a=1}^N 
  Z_a \delta(\mx{r}_{ia})   
+ \pi \sum_{i=1}^n \sum_{j>i}^n \delta(\mx{r}_{ij}) \nonumber \\ 
-&
\frac{1}{2} 
\sum_{i=1}^n\sum_{j>i}^n \left(%
   \frac{1}{r_{ij}} \mx{p}_i\cdot\mx{p}_j
 + \frac{1}{r_{ij}^3} \mx{r}_{ij}(\mx{r}_{ij}\cdot\mx{p}_i)\cdot\mx{p}_j
\right) \ .
\end{align}
%
In order to assess the uncertainty of the computations (Table~\ref{tab:err}), we evaluated the expectation values `directly' for the 
$\mx{p}_i^4$ and
$\pi\delta(\mx{r}_{ix})=\frac{1}{4}\nabla^2_{\mx{r}_{ix}}\frac{1}{r_{ix}} (x=j\ \text{or}\ a)$
operators \cite{StPaAd16} and by 
using the integral-transformation technique (IT) \cite{PaCeKo05}. 
Since we have accurate electronic wave functions, we expect that the two routes give very similar rovibrational intervals, still the results obtained with the IT integrals are expected to have a lower uncertainty.

The spin-independent $\alpha^3$-order QED corrections to the adiabatic potential energy of 
a diatomic molecule is \cite{BeSabook77,Pac98,PaKo10}: 
\begin{equation}\label{se3}
\begin{array}{@{}l}\displaystyle
E_{\rm rad}^{(3)} 
=
\alpha^3\frac{4}{3}\sum_{i=1}^n
    \left(
       \ln\frac{1}{\alpha^2}-\beta_{\rm el}+\frac{19}{30}
    \right)
    \left\langle 
       \phi_0 |Z\delta(\mathbf{r}_{i1})\!+\!Z\delta(\mathbf{r}_{i2}) |\phi_0
    \right\rangle
\\[3.5mm]\hspace{12mm}\displaystyle
    +\alpha^3 \sum_{i=1}^n \sum_{j>i}^n
    \left[
       \left(
          \frac{14}{3}\ln\alpha+\frac{164}{15}
       \right)
       \bigl\langle \phi_0 |\delta(\mathbf{r}_{ij})| \phi_0 \bigr\rangle
       -\frac{14}{3}Q_\text{el}
    \right],
\end{array}
\end{equation}
where
\begin{equation}\label{Bethe}
\beta_{\rm el} =
   \frac{
   \left\langle
      \phi_0|
      \mathbf{J}(H_0\!-\!E_0)\ln\left(2(H_0\!-\!E_0)/\mathrm{E}_\mathrm{h}
      \right)\mathbf{J}
      \phi_0      
   \right\rangle}
   {\left\langle
      \phi_0 |
      [\mathbf{J},[H_0,\,\mathbf{J}]]/2
   \phi_0
   \right\rangle}
\end{equation}
is the (non-relativistic) Bethe logarithm including the 
$\mathbf{J}\!=\!-\!\sum_{i=1}^n\mathbf{p}_i$ electric current density.
A precise evaluation of $\beta_\text{el}$ is a major numerical task and
values can be obtained if the wave function satisfies the electron-nucleus cusp condition \cite{BuJeMoKo92,Korobov_H2p_2013}. 
The $Q_\text{el}$ term was introduced by Araki and Sucher \cite{Ar57,Su58}:
\begin{equation}\label{Qterm}
  Q_\text{el} 
  = 
  \lim_{\epsilon \to 0} 
  \left\langle
    \phi_0 \Big| 
      \left[%
      \frac{\Theta(r_{ij} - \epsilon)}{ 4\pi r_{ij}^3 }
      + (\ln \epsilon + \gamma_E)\delta(\mathbf{r}_{ij}) 
      \right]
    \phi_0
  \right\rangle
\end{equation}
that is evaluated for He$_2^+$ using
the integral transformation technique \cite{PaCeKo05}
and the fECG basis representation.

Concerning the Bethe logarithm, we start with a few numerical observations.
Table~\ref{Bethe1} presents a compilation of the Bethe logarithm values for the lightest atoms and ions \cite{GoldmanDrake99,Pachucki_He10,Korobov_Bethe_He,Pachucki_Li,DrakeYan08} 
to highlight the weak dependence of $\beta_\text{el}$ on the number of electrons,
but its strong dependence on the nuclear charge, $Z$.
A similar observation applies for molecules described within the adiabatic approximation.
Table~\ref{Bethe2} shows the value of $\beta_{\rm el}(\rho)$ in the ground electronic state of 
the one-electron H$_2^+$ molecular ion and the two-electron H$_2$ molecule 
for selected values of the $\rho$ internuclear distance.
The $\beta_{\rm el}(\rho)$ values of H$_2^+$ and H$_2$ differ in the 4-5th significant digit.

These observations suggest that the Bethe logarithm of He$_2^+$ ($X\ ^2\Sigma_\text{u}^+$) 
can be well approximated with the Bethe logarithm of the ground electronic state of He$_2^{3+}$.
The Bethe logarithm for this one-electron two-center problem
was computed using the procedure of Ref.~\cite{Korobov_H2p_2013}.
We estimate the error introduced by the 
$\beta_\text{el,He$_2^+$}(\rho)\approx\beta_\text{el,He$_2^{3+}$}(\rho)$ 
approximation, which we use in this work,
to be less than $1$~\%\ over the relevant internuclear range, $\rho\in[0.9,3.5]$~bohr (Table~\ref{tab:err}).

\begin{table}
\caption{%
  Dependence of the $\beta_\text{el}$ Bethe logarithm on the $Z$ nuclear charge and 
  on the $n$ number of electrons in the ground state of atoms (ions).
  The data is compiled from Refs.~\cite{GoldmanDrake99,Pachucki_He10,Korobov_Bethe_He,Pachucki_Li,DrakeYan08}.}\label{Bethe1}
\begin{tabular}{@{}l D{.}{.}{8}  D{.}{.}{8}  D{.}{.}{8} @{}}
\cline{1-4}\\[-0.4cm]
\cline{1-4}\\[-0.4cm]
 & 
\multicolumn{1}{c}{\text{H}} &  
\multicolumn{1}{c}{\text{He}} & 
\multicolumn{1}{c}{\text{Li}} \\
$\beta_\text{el}$ [\Eh] & 
\multicolumn{1}{c}{$Z=1$} &  
\multicolumn{1}{c}{$Z=2$} & 
\multicolumn{1}{c}{$Z=3$} \\
\cline{1-4}\\[-0.4cm]
$n=1$ & 2.984128   & 4.370422 & 5.181353 \\
$n=2$ & \multicolumn{1}{c}{--} & 4.370160 & 5.179849 \\
$n=3$ & \multicolumn{1}{c}{--} & \multicolumn{1}{c}{--} & 5.17828  \\
\cline{1-4}\\[-0.4cm]
\cline{1-4}\\[-0.4cm]
\end{tabular}
\end{table}

\begin{table}
\caption{%
  Comparison of the $\beta_\text{el}(\rho)$ Bethe logarithm for selected $\rho$ internuclear distances
  of the one-electron H$_2^+$ molecular ion \cite{Korobov_H2p_2013} and 
  the two-electron H$_2$ molecule \cite{Pachucki_H2_2009}
  in the adiabatic approximation and in their ground electronic states. 
\label{Bethe2}}
\begin{tabular}{@{}l D{.}{.}{7} D{.}{.}{7} D{.}{.}{7} D{.}{.}{7} D{.}{.}{7} D{.}{.}{7} @{}}
\cline{1-7}\\[-0.4cm]\cline{1-7}\\[-0.4cm]
$\rho$ [bohr] & 0.1 & 0.2 & 0.4 & 0.8 & 1.5 & 5.0 \\
\cline{1-7}\\[-0.4cm]
$\beta_\text{el}(\rho)$(H$_2^+$) [\Eh] \cite{Korobov_H2p_2013} &
3.763208 &
3.525245 &
3.284256 &
3.100639 &
3.023053 &
2.995328 \\
$\beta_\text{el}(\rho)$(H$_2$) [\Eh] \cite{Pachucki_H2_2009} &
3.765 &
3.526 &
3.279 &
3.09331 &
3.01396 &
2.98534 \\
\cline{1-7}\\[-0.4cm]\cline{1-7}\\[-0.4cm]
\end{tabular}
\end{table}

The effect of higher-order QED corrections is estimated as in Refs.~\cite{Pachucki_H2_2009,PuKoCzPa16}:
\begin{align}
  E^{(4)}_\text{est}
  =
  \alpha^4\pi
  \left(%
    \frac{427}{96}-2\ln 2
  \right)
  \sum_{i=1}^3 \sum_{a=1}^2 Z_a \delta(\mx{r}_{ia}) \;.
  \label{eq:hQED}
\end{align}

Table~\ref{tab:err} collects the numerical uncertainty attributed to the rovibrational intervals 
within the described computational procedure.
The present theoretical framework rests on two small parameters, the square root of the electron-to-nucleus mass ratio, $\epsi$, and the fine-structure constant, $\alpha$. 
The electron-nucleus (non-adiabatic) coupling is accounted for up to $\epsi^2$ order and
higher-order contributions are neglected. 
Relativistic ($\alpha^2$) and leading-order QED ($\alpha^3$) corrections have been included, 
and an estimate for the $\alpha^4$-order terms, Eq.~(\ref{eq:hQED}),
was also computed. We estimate the uncertainty of the rotational-vibrational 
intervals due to the missing part of $\alpha^4$ and higher-order
QED corrections by the (small) effect of the $\alpha^4$ estimate (hQED in Table~\ref{tab:err}).
We have neglected the non-adiabatic-relativistic (and QED) coupling in the present work
that was found to be important in the H$_2$ molecule \cite{CzPuKoPa18}. 
An elaborate theoretical and computational study of 
this coupling for the present system will require further work, but
we give an estimate for its magnitude
(Nad\&Rel in Table~\ref{tab:err}). 
The estimated effect of the finite nuclear size is also shown in Table~\ref{tab:err}.
We used the CODATA18 recommendations for the 
physical constants and conversion factors throughout the computations. 

\begin{figure}
  \includegraphics[scale=0.82]{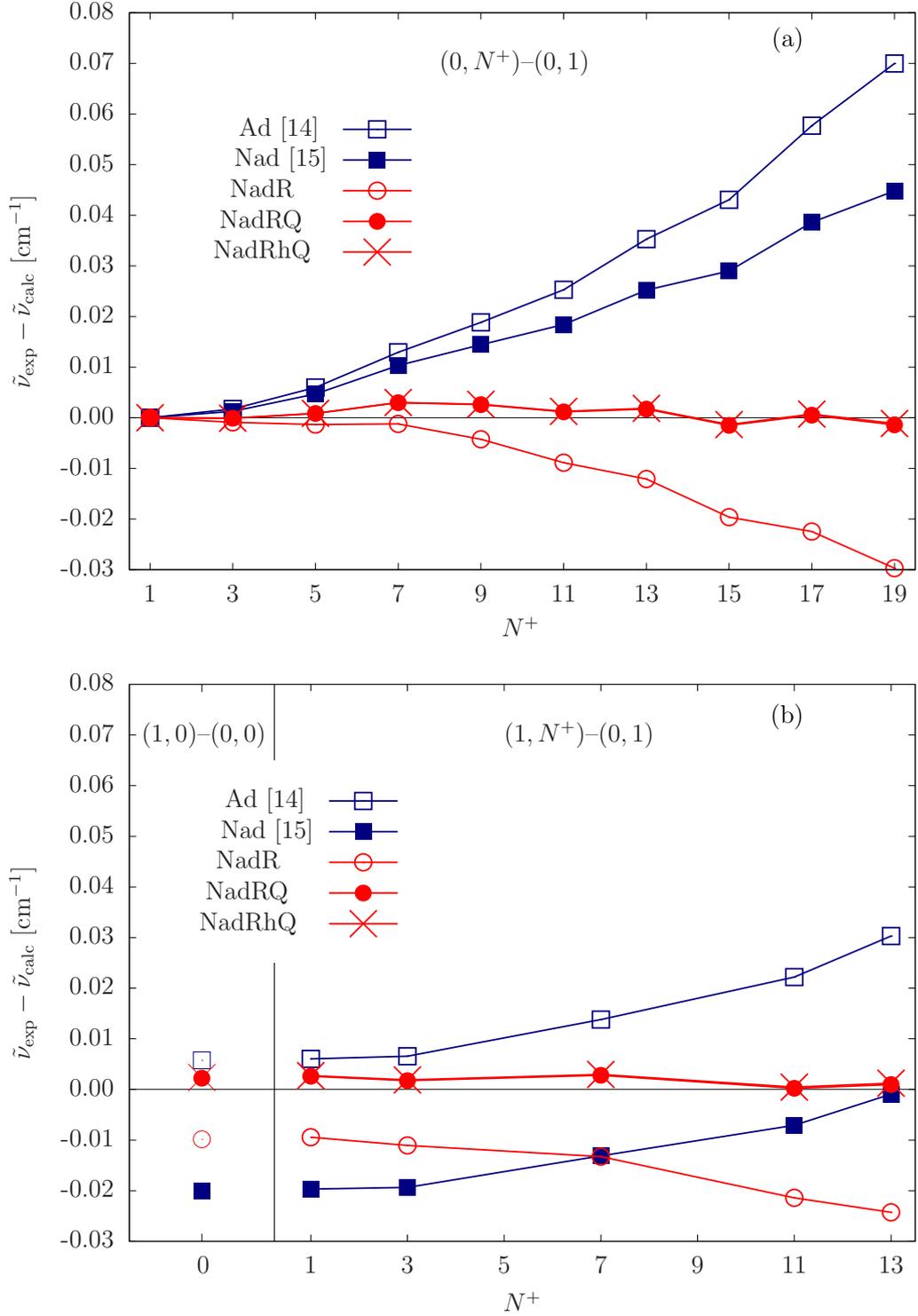}
  \caption{%
    Deviation of the rotational and (ro)vibrational excitation energies
    of the $^4$He$_2^+$ molecular ion ($X\ ^2\Sigma_\text{u}^+$) from experiment \cite{SeJaMe16,JaSeMe18,SeJaCaMeScMe20}.
    $(v,N^+)$ labels the $v$th vibrational state with the $N^+$ rotational quantum number. 
    {\color{blue}$\Box$}: adiabatic with 
    the $m_\text{vib}=m_\text{rot}=m_\alpha+1.5m_\text{e}$ empirical mass 
    in the kinetic energy operator; 
    {\color{blue}$\blacksquare$}: non-adiabatic, 
    \emph{i.e.,} with rigorous rotational and 
    vibrational masses; 
    {\Large\colred $\circ$}: non-adiabatic and relativistic corrections; 
    {\Large\colred $\bullet$}: non-adiabatic, relativistic, and leading-order QED corrections; 
    {\colred $\times$}: non-adiabatic, relativistic, leading-order QED with estimates for higher-order QED corrections.
    \label{fig:rovib}
  }
\end{figure}

\vspace{0.25cm}
The computed rotational and (ro)vibrational intervals and corrections are listed in Tables~\ref{tab:rot} and \ref{tab:rovib}.
Figure~\ref{fig:rovib} visualizes the results and 
reveals a fine interplay of the various corrections.
(The potential energy points and all corrections computed and used in this work are deposited in the Supplementary Material.)

%
The adiabatic description ('Ad') with the `empirical mass correction' 
using $m_\text{rot}=m_\text{vib}=m_\alpha+1.5m_\text{e}$ \cite{TuPaAd12}
reproduces the fundamental vibration energy 
almost perfect, while its deviation from experiment increases with increasing $N^+$.
By including the rigorous non-adiabatic masses for the rotational and vibrational degrees of freedom \cite{Ma18he2p}, 
the error is reduced for the rotational excitations, but the fundamental vibration energy shows a large deviation from experiment. 
Adding the relativistic corrections to this non-adiabatic model reduces the deviation 
to the half for the fundamental vibration, but it `over-corrects' the rotational excitation energies.
By including also the leading-order QED corrections in the theoretical treatment both the fundamental vibration
energy, the rotational and the rovibrational excitation energies come in agreement with experiment with 
a root-mean-squared deviation (RMSD) of  $0.001\,7$ and $0.001\,9$~\cm, respectively.
The experimental uncertainty of the rotational and rovibrational series is slightly smaller
than these values \cite{SeJaMe16,JaSeMe18}, they are $0.000\,8$ and $0.0001\,2$~\cm, respectively.
The lowest-energy rotational interval, $(0,3)$--$(0,1)$,
has been recently measured more precisely,
$70.937\,589(23)\pm 0.000\,06_\text{sys}$
\cite{SeJaCaMeScMe20}, and our 
theoretical value for this interval is 
$70.937\,69(10)$~\cm.
For the fundamental vibration, our computational result is 
$1628.380\,9(33)$~\cm, which is in agreement with its value derived from experiments,
$1628.383\,2(12)$~\cm\ \cite{JaSeMe18}.

All the rovibrational intervals (Table~\ref{tab:rovib}) are in agreement with the experimental results within the given uncertainties, although the computational results have 
almost three times larger uncertainties than the experimental ones.
We observe some discrepancy 
for the \emph{rotational} intervals with intermediate $N^+$ values (especially, $N^+=7,9$ and 13).
We note that the pure rotational intervals have a smaller uncertainty than the rovibrational ones, since they were much less affected by the PEC improvement (Table~\ref{tab:err}).

We finish the discussion with observations
regarding the interplay of the computed corrections (Tables~\ref{tab:rot} and \ref{tab:rovib}).
First, we point out that $\delta\tilde\nu_\text{mveff}$ and $\delta\tilde\nu_\text{mvNad}$
together account for the non-adiabatic mass effect. $\delta\tilde\nu_\text{mveff}$ is a simple,
intuitive, constant mass model ($m_\text{rot}=m_\text{vib}=m_\alpha+1.5m_\text{e}$)
and $\delta\tilde\nu_\text{mvNad}$ labels the value, which corrects this empirical model to
arrive at the rigorous second-order non-adiabatic result. 
It is interesting to observe, at least for the present example, 
that $\delta\tilde\nu_\text{mvNad}$ has the same order of magnitude but opposite sign
as the leading-order QED correction, $\delta\tilde\nu_\text{QED}$.
The interplay of the corrections changes for the different types of motions, \emph{i.e.,} the relativistic correction
has a different sign for the rotational and for the vibrational excitation, whereas the QED contribution is positive 
in both cases.
This interplay of the higher-order correction terms---that we explicitly compute in the present work---had resulted in cancellation of errors in the lower-order calculations \cite{TuPaAd12} and a seemingly good agreement with the experimental result \cite{JaSeMe18} for this interval.

\clearpage
\begin{table}
  \caption{%
    Rotational excitation energies of $^4$He$_2^+$ ($X\ ^2\Sigma_\text{u}^+$) in the vibrational ground state.
    $\tilde\nu_0$: Born--Oppenheimer description with nuclear masses; 
    $\delta\tilde\nu_\text{DBOC}$: the diagonal Born--Oppenheimer correction; 
    $\delta\tilde\nu_\text{mveff}$: empirical mass: $m_\text{rot}=m_\text{vib}=m_\alpha+1.5m_\text{e}$;
    $\delta\tilde\nu_\text{Nad}$: rigorous non-adiabatic mass;
    $\delta\tilde\nu_\text{Rel}$: relativistic correction;
    $\delta\tilde\nu_\text{QED}$: leading-order QED correction;
    $\delta\tilde\nu_\text{hQED}$: estimate for higher-order QED corrections. 
    For the derivation of the error estimates to the computed energies see Table~\ref{tab:err}.
\label{tab:rot}
}
\begin{flushleft}
\begin{tabular}{@{}l @{\ \ } D{.}{.}{16} @{\ \ \ } . @{\ \ \ \ }. @{\ \ \ \ }. @{\ \ \ \ }. @{}}
\cline{1-5}\\[-0.4cm] 
\cline{1-5}\\[-0.4cm]
 & \multicolumn{4}{c}{$\tilde\nu(0,N^+)-\tilde\nu(0,1)$ [\cm]} \\
\cline{2-5}\\[-0.4cm]
 \multicolumn{1}{c}{$N^+$:} &	
 \multicolumn{1}{c}{3~~~~~~~~~~~~~~~~~~~~~} &
 \multicolumn{1}{c}{5~~~~~~} &
 \multicolumn{1}{c}{7~~~~~~} &	
 \multicolumn{1}{c}{9~~~~~~} \\
\cline{1-5}\\[-0.4cm]

$\tilde\nu_0$                   &	70.96061	&	198.4278	&	381.9543	&	620.8981	\\
$+\delta\tilde\nu_\text{DBOC}$	& -0.01028	&	-0.0287	&	-0.0550	&	-0.0891	\\
$+\delta\tilde\nu_\text{mveff}$  &	-0.01446	&	-0.0404	&	-0.0776	&	-0.1258	\\
$+\delta\tilde\nu_\text{mvNad}$	&	0.00045	&	0.0013	&	0.0026	&	0.0044	 \\
$+\delta\tilde\nu_\text{Rel}$	&	0.00216	&	0.0060	&	0.0115	&	0.0187 \\
$+\delta\tilde\nu_\text{QED}$	&	-0.00078	&	-0.0022	&	-0.0042	&	-0.0068	\\
$+\delta\tilde\nu_\text{hQED}$	&	0.00001	&	0.0000	&	0.0000	&	-0.0001	\\
$\tilde\nu_\text{calc}$         &  70.93768(10)	&	198.3638(13)&	381.8316(13)&	620.6994(13) \\
\cline{1-5}\\[-0.4cm]
$\tilde\nu_\text{expt}$ \cite{SeJaCaMeScMe20,SeJaMe16} &	70.937589(23)(60)_\text{sys}   &	198.3647(8)	&	381.8346(8)	&	620.7021(9)	\\
$\tilde\nu_\text{expt}-\tilde\nu_\text{calc}$  &	-0.00010	&	0.0009	&	0.0030	&	0.0027 \\
\cline{1-5}\\[-0.4cm]
\end{tabular}

\begin{tabular}{@{}l ... .. @{}}
 & \multicolumn{5}{c}{$\tilde\nu(0,N^+)-\tilde\nu(0,1)$ [\cm]} \\
\cline{2-6}\\[-0.4cm]
 \multicolumn{1}{c}{$N^+$:} &	
 \multicolumn{1}{c}{11~~~~~~} &	
 \multicolumn{1}{c}{13~~~~~~} &
 \multicolumn{1}{c}{15~~~~~~} &
 \multicolumn{1}{c}{17~~~~~~} &
 \multicolumn{1}{c}{19~~~~~~} \\
\cline{1-6}\\[-0.4cm]
$\tilde\nu_0$                   &	914.4265	&	1261.5215	&	1660.9860	&	2111.4508	&	2611.3826	\\
$+\delta\tilde\nu_\text{DBOC}$	&	-0.1304	&	-0.1788	&	-0.2336	&	-0.2944	&	-0.3605	\\
$+\delta\tilde\nu_\text{mveff}$  &	-0.1847	&	-0.2538	&	-0.3328	&	-0.4209	&	-0.5177	\\
$+\delta\tilde\nu_\text{mvNad}$	&	0.0069	&	0.0101	&	0.0141	&	0.0191	&	0.0252	\\
$+\delta\tilde\nu_\text{Rel}$	&	0.0273	&	0.0373	&	0.0486	&	0.0611	&	0.0745	\\
$+\delta\tilde\nu_\text{QED}$	&	-0.0100	&	-0.0138	&	-0.0181	&	-0.0230	&	-0.0283	\\
$+\delta\tilde\nu_\text{hQED}$	&	-0.0001	&	-0.0001	&	-0.0002	&	-0.0002	&	-0.0003	\\
$\tilde\nu_\text{calc}$ &	914.1354(13) &	1261.1223(13) &	1660.4640(13)	&	2110.7925(13)	&	2610.5755(15) \\
\cline{1-6}\\[-0.4cm]
$\tilde\nu_\text{expt}$ \cite{SeJaMe16} &	914.1367(8)	&	1261.1242(8)	&	1660.4627(9)	&	2110.7932(9)	&	2610.5744(9)	\\
$\tilde\nu_\text{expt}-\tilde\nu_\text{calc}$ & 0.0013	&	0.0019	&	-0.0013	&	0.0007	&	-0.0011	\\
\cline{1-6}\\[-0.4cm]
\cline{1-6}\\[-0.4cm]
\end{tabular}
\end{flushleft}
\end{table}

\clearpage
\begin{table}
  \caption{%
    Rovibrational excitation energies of $^4$He$_2^+$ ($X\ ^2\Sigma_\text{u}^+$)
    between the vibrational ground and first excited state.
    See also the caption to Table~\ref{tab:rot}.
    \label{tab:rovib}
  }
\begin{tabular}{@{}l@{\ \ \ \ \ \ } .@{\ \ \ \ }.. @{}}
\cline{1-4}\\[-0.4cm] 
\cline{1-4}\\[-0.4cm]
 & \multicolumn{3}{c}{$\tilde\nu(v,N^+)''-\tilde\nu(v,N^+)'$ [\cm]} \\
\cline{2-4}\\[-0.4cm] 
 \multicolumn{1}{c}{$(v,N^+)''$--$(v,N^+)'$:} &
 \multicolumn{1}{l}{$(1,0)$--$(0,0)$} &
 \multicolumn{1}{l}{$(1,1)$--$(0,1)$} &
 \multicolumn{1}{l}{$(1,3)$--$(0,1)$} \\
\cline{1-4}\\[-0.4cm]
$\tilde\nu_0$        	        &	1628.5600	&	1628.1081	&	1696.8089 \\
$\delta\tilde\nu_\text{DBOC}$   &	-0.0223	&	-0.0222	&	-0.0320	\\
$\delta\tilde\nu_\text{mveff}$  &	-0.1602	&	-0.1601	&	-0.1739	\\
$\delta\tilde\nu_\text{mvNad}$  &	0.0258	&	0.0257	&	0.0259	\\
$\delta\tilde\nu_\text{Rel}$	&	-0.0102	&	-0.0103	&	-0.0083	\\
$\delta\tilde\nu_\text{QED}$	&	-0.0120	&	-0.0120	&	-0.0128	\\
$\delta\tilde\nu_\text{hQED}$	&	-0.0001	&	-0.0001	&	-0.0001	\\
$\tilde\nu_\text{calc}=\tilde\nu_0+\sum\delta\tilde\nu$ &	1628.3809(33)	&	1627.9291(39)	&	1696.6077(39) \\
\cline{1-4}\\[-0.4cm]
$\tilde\nu_\text{expt}$ \cite{JaSeMe18} &	1628.3832(12)	&	1627.9318(12)	&	1696.6096(12)	\\
$\tilde\nu_\text{expt}-\tilde\nu_\text{calc}$ &	0.0023	&	0.0027	&	0.0019	\\
\cline{1-4}\\[-0.4cm]
%
%
%
 & \multicolumn{3}{c}{$\tilde\nu(v,N^+)''-\tilde\nu(v,N^+)'$ [\cm]} \\
\cline{2-4}\\[-0.4cm] 
 \multicolumn{1}{c}{$(v,N^+)''$--$(v,N^+)'$:} &
 \multicolumn{1}{l}{$(1,7)$--$(0,1)$} &
 \multicolumn{1}{l}{$(1,11)$--$(0,1)$} &
 \multicolumn{1}{l}{$(1,13)$--$(0,1)$} \\
\cline{1-4}\\[-0.4cm]
$\tilde\nu_0$        	        &	1997.8578	&	2513.1465	&	2848.9316	\\
$\delta\tilde\nu_\text{DBOC}$   &	-0.0744	&	-0.1459	&	-0.1916	\\
$\delta\tilde\nu_\text{mveff}$  &	-0.2339	&	-0.3357	&	-0.4013	\\
$\delta\tilde\nu_\text{mvNad}$  &	0.0269	&	0.0293	&	0.0312	\\
$\delta\tilde\nu_\text{Rel}$	&	0.0002	&	0.0143	&	0.0233	\\
$\delta\tilde\nu_\text{QED}$	&	-0.0161	&	-0.0216	&	-0.0252	\\
$\delta\tilde\nu_\text{hQED}$	&	-0.0002	&	-0.0002	&	-0.0003	\\
$\tilde\nu_\text{calc}=\tilde\nu_0+\sum\delta\tilde\nu$ &	1997.5604(39)	&	2512.6867(39)	&	2848.3678(39)	\\
\cline{1-4}\\[-0.4cm]
$\tilde\nu_\text{expt}$ \cite{JaSeMe18} &	1997.5633(12)	&	2512.6871(12)	&	2848.3690(12) \\
$\tilde\nu_\text{expt}-\tilde\nu_\text{calc}$ &	0.0029	&	0.0004	&	0.0012	\\
\cline{1-4}\\[-0.4cm]
\cline{1-4}\\[-0.4cm]
\end{tabular}
\end{table}

\clearpage
Rotational and (ro)vibrational intervals have been reported for 
the three-electron $^4$He$_2^+$ ($X\ ^2\Sigma_\text{u}^+$) molecular ion 
on a newly computed potential energy curve with non-adiabatic,
relativistic, and QED corrections. 
The computed rotational-vibrational intervals are 
in good agreement with recent precision spectroscopy measurements. 
Further developments, most importantly, 
a detailed study of the relativistic-non-adiabatic coupling
and the extension of the potential energy curve with ppb uncertainty over large 
internuclear distances, will challenge precision spectroscopy experiments 
and contribute to the establishment of primary pressure standards.

\begin{acknowledgments}
DF acknowledges the support from the ÚNKP-19-3 New National Excellence Program of the Ministry for Innovation and
Technology (ÚNKP-19-3-I-ELTE-24). 
EM and DF acknowledge the financial support of the Swiss National Science Foundation 
(PROMYS Grant, No.~IZ11Z0\_166525) at the beginning of this work
and the European Research Council (Starting Grant, No.~851421).
EM thanks Frédéric Merkt for discussions.
V.I.K. acknowledges support from the Russian Foundation for Basic Research under Grant No. 19-02-00058-a.
\end{acknowledgments}


%

\end{document}